\documentstyle[epsf,prb,aps,eqsecnum]{revtex}
\tighten
\begin{document}
\draft
\title{Mean-Field Nematic--Smectic-{\sl A} Transition in a Random
Polymer Network}
\author{Peter D. Olmsted\footnote{e-mail:
{\it pdo@calvino.physics.lsa.umich.edu};
present address, Dept. of Physics, University of Michigan, Ann Arbor, MI
48109-1120; permanent address, Dept. of Physics, University of Leeds,
Leeds LS2 9JT, UK}
and Eugene M.~Terentjev}
\address{ Cavendish Laboratory,
Madingley Road, Cambridge,   CB3 0HE, U.K.}
\date{\today}
\maketitle

\begin{abstract}
Liquid crystal elastomers present a rich combination of effects
associated with orientational symmetry breaking and the underlying
rubber elasticity. In this work we focus on the effect of the network
on the nematic--smectic-{\sl A} transition, exploring the additional
translational symmetry breaking in these elastomers. We incorporate the
crosslinks as a random field in a microscopic picture, thus expressing
the  degree to which the smectic order is locally frozen with respect
to the network.  We predict a modification of the NA transition,
notably that it can be treated at the mean-field level (type-I system),
due to the coupling with elastic degrees of freedom. There is a shift
in the transition temperature $T_{\scriptscriptstyle NA}$, a
suppression of the Halperin-Lubensky-Ma (HLM) effect (thus recovering
the mean-field continuous transition to the smectic state), and a new
tri-critical point, depending on the conditions of network formation.
When the nematic phase possesses `soft elasticity', the NA transition
becomes of first order due to the coupling with soft phonons in the
network.  We also discuss the microscopic origin of phenomenological
long-wavelength coupling between smectic phase and elastic strain.
\end{abstract}

\widetext

\section{Introduction}
Randomly crosslinked networks of polymer liquid crystal  (LCP)
materials (liquid crystalline elastomers and gels) have been a subject
of substantial experimental and theoretical activity in recent years. A
newcomer to this area can find out about about  the synthesis of
side-chain and main-chain systems, characteristic physical effects, and
concepts related to the nematic state in  review articles
\cite{barclayober93,warter95}. Three major factors determine the
behavior of these remarkable materials: liquid crystalline symmetry
breaking; rubber elasticity coupled to the resulting anisotropy,
producing a highly mobile principal axis; and, finally, randomly placed
(and sometimes randomly oriented) network crosslinks.  Random disorder
introduced by these crosslinks encourages elastomers formed in the
isotropic state to cool into polydomains, {\it i.e.\/} highly
non-uniform textures of their liquid crystalline phases. If the network
is formed in the presence of an external field, or in a uniform
monodomain low-temperature phase, the resulting elastomer remains
macroscopically uniform.

Smectic elastomers (refs. \cite{symons93,brehmer94,finkelmann94}
represent a few recent examples) comprise liquid crystalline
polymers crosslinked into a network which contains liquid crystalline
groups ordered into a one-dimensional density wave, or a smectic
state.  One may consider either main chain or, more commonly,
side-chain LCP's.  The smectic order is typically obtained by forming
the network in a smectic state ({\it e.g.\/} by crosslinking at
temperatures below the nematic--smectic-{\sl A}, or NA, transition
temperature $T^0_{\scriptscriptstyle\it NA}$ in the melt). It can  also
be obtained by cooling a sample that has been prepared in the nematic
state through the transition temperature $T_{\scriptscriptstyle\it
NA}$, where we expect $T_{\scriptscriptstyle\it NA} <
T^0_{\scriptscriptstyle\it NA}$.  The NA transition in networks is
expected to be qualitatively very different from the same transition in
melts because of the presence of elastic strain degrees of freedom and
the local effects of random crosslinks.

In the conventional NA transition, the degrees of freedom are the smectic
order parameter $\psi({\rm\bf r})$ and the director fluctuations $\delta
\bbox{n}$.  The parameter $\psi({\rm\bf r})$ describes the departure of the
mesogen center-of-mass density $\rho({\rm\bf r})$ from a uniform density
$\rho_0$, in a form of single wavelength modulation  \cite{tclchim83,pgdg}
\begin{equation}
\rho({\rm\bf r}) = \rho_0\left\{ 1 + {1\over\sqrt{2}}\left[\psi({\rm\bf r})
e^{i {\rm\bf q}_0\cdot{\rm\bf r}} + {\it c.c}\right]\right\}.
\end{equation}
The nematic state is parameterized by the director $\bbox{\hat{n}}$,
which indicates the axis of preferential alignment of the mesogenic
groups,  and $\delta\bbox{\hat{n}}$ indicates the fluctuations
of this axis. While the NA transition should be continuous according to
symmetry arguments, director fluctuations may change this picture.
Halperin, Lubensky, and Ma \cite{hlm} argued that the coupling
to the fluctuating ``gauge'' field $\delta \bbox{n}$ should induce a
first order transition in type-I smectics (and the analogous
superconducting system). In type-I smectics the characteristic length for
penetration of the  director twist into the ordered smectic state is much
smaller that the correlation length for the smectic order parameter, so
that director fluctuations may be treated at a mean field level. Type-II
smectics and superconductors, in which the gauge field
fluctuations are correlated on length scales comparable or larger
than that of the order parameter $\psi ({\rm\bf r}) $,  require a
more sophisticated analysis.
The beliefs about the nature of this transition have varied over the
years. A renormalization group $\epsilon$-expansion \cite{hlm} yielded no
fixed point for physical order parameters, suggesting that the transition
is weakly first order. However, duality arguments combined with Monte
Carlo calculations on a lattice \cite{dasgupta81} predicted a continuous
transition in the universality class of the inverted 3D XY-transition
(that is, the amplitude ratios are inverted), experiments have generally
yielded a continuous transition of the non-inverted 3D XY class
\cite{bouwman92,garland93I,garland93II} (but see \cite{anisimov90}).
Toner \cite{toner82} has argued, on the basis of a dislocation-melting
theory, that the transition should be continuous; Andereck and Patton
\cite{andereck94} have calculated critical exponents within
self-consistent perturbation theory; and Radzihovsky \cite{radz95} has
performed a similar calculation for the analogous type-II superconducting
transition, demonstrating the existence of a fixed point not found in
$\epsilon$-expansion of \cite{hlm}.

While our understanding of the NA transition could be
incomplete, the nature of
the smectic state is well-understood. The smectic one-dimensional density
wave, of wave number $q_0$, suffers the Landau-Peierls instability
and the system exhibits quasi long ranged order
(with an algebraic decay of density correlations at long distances)
\cite{caille,als80}. The source of this instability is the
Goldstone modes corresponding to the arbitrary phase of the smectic
density wave, whose long wavelength free energy is given by the
Landau-Peierls  elastic energy of a one-dimensional solid \cite{pgdg69}.
In three dimensions this energy yields algebraic decay of
correlations in the ordered state, analogous to systems with broken
continuous symmetry in two dimensions.

In contrast to a liquid crystalline melt, a liquid crystalline
elastomer has an additional set of long-wavelength degrees of freedom:
the elastic deformation ${\rm\bf v}({\rm\bf r})$, defined by the local
network displacement, ${\rm\bf r}\longrightarrow{\rm\bf r} + {\rm\bf v}({\rm\bf
r})$.
Smectic order couples to the elasticity through the crosslinks,
and this coupling is manifested in two effects:

\begin{itemize}
\itemindent=-1.0em
\item[A.] Crosslinks pin the smectic phase and
break translational invariance.
The mesogens are either incorporated into the
polymer backbone (main chain LCP's) or tethered to the
network (side chain LCP's), and the stretching energy for relative
translations between elastic displacements of the crosslinks
$v_z({\rm\bf r})$ along the direction
of the smectic density wave and the layer displacement $u({\rm\bf r})$
can be written phenomenologically as \cite{terwarlub95}
\begin{equation}
2 F_{d} = \Lambda\int\!d^3\!r [v_z({\bf r }) -  u({\bf r })]^2.
\label{eq:new}
\end{equation}
This energy is defined on lengthscales of order or longer than the
characteristic mesh size and, most importantly, is present for uniform
relative translations. This coupling restores the Bragg peaks
associated with smectic order \cite{terwarlub95}.  In Section~II we
argue that this term is in fact only a {\it metastable\/} term, and the
relative displacement $v_z({\bf r }) -  u({\bf r })$ may be relaxed by
layer hopping \cite{renz86} with a characteristic relaxation time.

\item[B.] The preferential reduction of mesogen density
around a crosslink due to, {\it e.g.\/}, the steric exclusion of the mesogen,
leads us (see Section~II) to model the crosslinks by a local random
field which adjusts the smectic phase:
\begin{equation}
F_{\scriptscriptstyle RF} = \gamma\int\!d^3\!r\,
c({\rm\bf r})|\psi({\rm\bf r})|
\cos\left[q_0(z - u({\rm\bf r}) + v_z({\rm\bf r}))\right],
\label{eq:fxl0}
\end{equation}
where $c({\rm\bf r})$ is the local crosslink concentration.
\end{itemize}

The smectic elastomer thus contains four contributions to
the continuum free energy in addition to the smectic and
nematic free energies associated with the NA transition in liquids. These are:

\begin{itemize}
\itemindent=-1.0em
\item[(i)] The elastic free energy of a uniaxial solid
\cite{landauelastic},  written in terms of the
symmetric elastic strain $\epsilon_{\alpha\beta} =
\case12(\nabla_{\alpha}v_{\beta} + \nabla_{\beta}v_{\alpha})$.
This contribution describes the phonon field (in most elastomers
- incompressible) in the rubbery network. It couples to the
relevant variables of the NA transition through the following
effects.

\item[(ii)] Eq.~(\ref{eq:new}), the term penalizing
relative shifts in the smectic phase variable $u({\rm\bf r})$ and
the phonon displacement parallel to it, $v_z({\rm\bf r})$.

\item[(iii)] Rubber-nematic couplings \cite{pgdg80one},
penalizing relative rotations of the elastic
strain $\bbox{\Omega}=\case12(\nabla\times{\rm\bf v})$ and the
nematic director,
$\bbox{\omega}=\bbox{\hat{n}}\times\delta\bbox{\hat{n}}$.

\item[(iv)] The random field term, Eq.~(\ref{eq:fxl0}),
describing the effect of crosslinks on the phase of the smectic
order parameter.
\end{itemize}

In this work we consider the microscopic origin of Eq.~(\ref{eq:new})
and treat the random field in Eq~(\ref{eq:fxl0}) by the replica formalism
\cite{edwardsanderson75}, with the following results [embodied in
Eqs.~(\ref{eq:Flandboth})] for the mean-field theory
of the NA transition in the network:

\begin{itemize}
\itemindent=-1.0em
\item[1.] Near the putative continuous NA transition
Eq.~(\ref{eq:new}) may be ignored, since it is primarily a dynamic
effect. At low enough temperatures (when the characteristic timescale
becomes essentially infinite) this term must be considered.

\item[2.] At the mean field level nematic director fluctuations no
longer induce the Halperin-Lubensky-Ma first order smectic phase
transition: they acquire a mass which reduces their effect on the NA
transition. The existence of this mass in fact makes the mean field
treatment (and the associated type-I assumption) essentially exact.

\item[3.] However, in the special case where the nematic elastomer
possesses {\sl soft elasticity\/} (see Eq.~\ref{eq:csoft} below),
characteristic of spontaneously breaking the symmetry from the
isotropic to nematic states \cite{golubovic89,olmsted94}, elastic
strain fluctuations [the phonons ${\rm\bf v}({\rm\bf r})$] restore the
HLM effect and the concomitant first order phase transition.

\item[4.] If the nematic state is field-induced rather than
spontaneous ({\it i.e.\/} the network has been formed to record the
broken orientational symmetry), the corresponding network phonon modes
are conventional and the type-I NA transition should be continuous.  We
predict a tri-critical point in the crossover region between these two
regimes.

\item[5.] The effect of disorder on the NA transition is a simple
renormalization of the transition temperature $T_{\scriptscriptstyle
NA}$.  The transition temperature {\sl increases\/} because
crosslinks localize the smectic phase variable and encourage order in
the disordered phase. However, slightly below $T_{\scriptscriptstyle
NA}$ the smectic state crosses over to a `glassy' state characterized
by replica--symmetry-breaking.

\item[6.] We emphasize that the typical nematic or smectic network
is most likely to be `hard', and that the `soft' case we discuss
extensively here is, for most practical cases, a theoretical construct
which provides a framework for understanding the implications of a
smectic state coupled to an underlying elastic continuum.
\end{itemize}

Our description of the NA transition in the network is
mesoscopic because we treat the smectic and director degrees of freedom
as smoothly varying fields while including
the microscopic interaction of the crosslinks with the smectic
order parameter. The calculation in this paper sets up the proper
coarse-grained model with which to investigate the effects of disorder
on the smectic network, which we discuss briefly here and
in more detail in a forthcoming work. The coarse-grained
nematic and smectic fields are defined on a length $\ell_{\it lc}$ of
order a correlation length for liquid-crystalline fluctuations, while
the elastic strain field is coarse-grained on lengths down to the mesh
size $\ell_{\it x}$ in the network, which may be larger than $\ell_{\it
lc}$.  These different lengthscales must be kept in mind
throughout.

This paper is organized as follows: in Section~II we present a more
detailed discussion of the two primary effects of the network-smectic
couplings; in Section~III we present the model free energy for the
system;
in Section~IV we sketch the integration of the elastic strain
and director fluctuations and present an effective free energy for the
smectic in the presence of disorder; and we conclude in Section~V. Our
primary results are the effective Landau energies at the end of
Section~IV and the discussion in Section~II about the smectic-elastic
couplings.  Our goal here is to understand the various terms in the
Hamiltonian describing the NA transition, and identify the effects of
disorder and rubber elasticity on the instability to the smectic state.
We leave for the future a detailed analysis of the effects of disorder
on the low temperature smectic state.  The appendix contain some
technical details of the replica calculations.

\section{Smectic--Elastic Couplings}
\subsection{Random Field}
In this section we justify the random field coupling,
Eq.~(\ref{eq:fxl0}). This interaction represents the pinning of the
smectic  phase to network inhomogeneities, which we represent by a
random distribution of crosslinks. Let us focus on side-chain LCP's,
Fig.~1,
since these constitute the most commonly
synthesized liquid crystalline elastomers.  The three
constituents of the elastomer are flexible backbone monomers, crosslink
groups, and mesogenic side-groups attached to the backbone with
flexible spacers. In elastomers (as opposed to tough and brittle
densely-crosslinked resins) the volume fraction of crosslinks is much
smaller than that of the backbone and mesogenic groups, typically less than a
percent of the aggregate. Hence we treat the environment of the
mesogens as a uniform mesogen/backbone mixture with {\it
interspersed\/} crosslink groups.
\begin{figure}
\epsfxsize=3truein
\centerline{\epsfbox{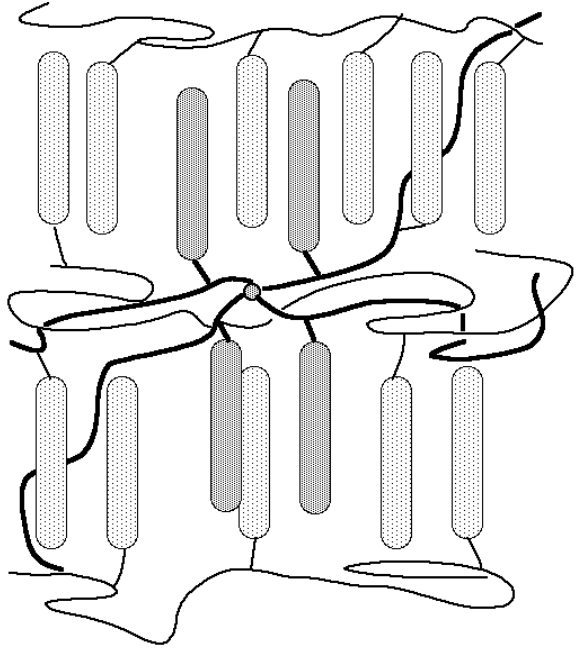}}
\caption{Sketch of mesogenic groups in a smectic elastomer.}
\label{fig1}
\end{figure}
It is reasonable to expect a steric
repulsion between the mesogen and the crosslink which is enhanced
relative to the steric repulsion between the backbone and the mesogen,
simply because four chains come together at the crosslink. The presence
of a Flory-$\chi$ parameter between the mesogen and backbone groups
contributes a similar effect. While there
will also be an interaction with the local nematic order, since the
mesogens may adopt an alignment dictated by the crosslink, we
concentrate in this work on the effects of disorder on a smectic phase
appearing out of a uniform conventional nematic state. Hence, we model
the crosslinks by a local random field which induces smectic order and
fixes the smectic phase, leading to Eq.~(\ref{eq:fxl0}). The
corresponding energy contribution for each crosslink in the smectic
potential is
\begin{equation}
F_{\scriptscriptstyle RF} = \sum_{i} \gamma
|\psi({\rm\bf R}_i)|
\cos \left[ q_0 (z_i - u({\rm\bf R}_i))\right] \ ,
\label{eq:fxlsum}
\end{equation}
where ${\rm\bf R}_i$ is the position of the $i^{\rm th}$ crosslink.  The
coupling constant $\gamma$ can be estimated on the basis of Fig.~1.  We
shall assume that the crosslinking point localizes the positions of
$\phi$ monomers (with $\phi$ the crosslink functionality). The barrier
for such an object to `tunnel' through a smectic smectic layer is a
molecular characteristic of the material describing the degree of
miscibility of backbone and the mesogenic side-groups, and is roughly
of order $\phi\chi$, where $\chi$ is the the Flory-$\chi$ parameter
between mesogens and backbone, which includes both steric and energetic
interactions.

Introducing the continuum crosslink density
$c({\rm\bf r}) = \sum_i\delta({\rm\bf r}-{\rm\bf R}_i) $, so that
under an elastic distortion ${\rm\bf R}_i\rightarrow{\rm\bf R}_i+{\rm\bf v}$
they distort by
$c({\rm\bf r})\rightarrow c({\rm\bf r} - {\rm\bf v})$, we can transform
(\ref{eq:fxlsum}) to collective variables. After changing
variables ${\rm\bf r}' = {\rm\bf r} - {\rm\bf v}({\rm\bf r})$ we obtain
\begin{eqnarray}
F_{\scriptscriptstyle RF} &=& \gamma \int\!d^3\!r\, c({\rm\bf r})\,
|\psi({\rm\bf r} + {\rm\bf v}({\rm\bf r}))|
\cos\left[q_0\left(z - u({\rm\bf r}) + v_z({\rm\bf r})\right)\right].
\label{eq:fxl}
\end{eqnarray}
(Changing variables in the argument of $u({\rm\bf r})$
introduces higher-order gradient corrections which are
irrelevant in a mean-field treatment).

Rather than working with the discrete
crosslink positions, we represent the crosslink concentration
$c({\rm\bf r})$ by a Gaussian probability distribution
\cite{edwardsmuthukumar88}:
\begin{equation}
P[c]\propto \exp\left\{-\frac{1}{2}\int\!d^3\!r \, \frac{c({\rm\bf r})^2}
{2 N_x}\right\} \ ,   \label{eq:Pc}
\end{equation}
where $N_x$ is number of
crosslinks per unit volume of the system. For this
random distribution of crosslinks the characteristic moments are
$\langle c({\rm\bf r}) \rangle = N_x $ \ and \
$\langle c({\rm\bf r}) c({\rm\bf r}') \rangle = N_x \delta
({\rm\bf r}-{\rm\bf r}')$.

One should note, at this point, that a different situation emerges for
a network formed in a deep uniform smectic phase: the polymer backbone
in such a system is highly constrained between the smectic layers and,
therefore, the crosslink distribution  (\ref{eq:Pc}) has a positionally
modulated kernel. This possibility will have a profound effect on the
phase ordering in the smectic phase, but is not relevant for the NA
transition description we are concerned here. Also, in a physical system
crosslinks fluctuate about their mean positions \cite{flory77}, which
broadens the kernel of Eq.(\ref{eq:Pc}) into a Gaussian distribution
with a width (in real space) proportional to the extent of the typical
crosslink fluctuation. We ignore such fluctuation effects for now.

\subsection{Translational Invariance}
The second effect of the network on the smectic phase is due to
stretching the polymer backbone. This leads to a free energy cost
for uniformly displacing the smectic phase relative to the elastic
displacement, given by Eq.(\ref{eq:new}) \cite{terwarlub95}.
Whereas the contribution of the previous sub-section acts only at the
crosslink position, the stretching effect influences the smectic phase
all along the strand between crosslinks, and is coarse-grained at the mesh
size or larger.  To understand this in detail we first note the distinction
between the smectic phase variable $q_0u$ and the $z$-displacement
$\eta_i$ of the center-of-mass of the $i^{th}$ smectic mesogen.

The phase $q_0u$ is a coarse-grained variable and the displacements
$\{\eta_i\}$ are microscopic variables. To calculate the position of
the smectic phase from a microscopic picture one must, in principle,
average over all the positions of the mesogens. In an equilibrium
smectic {\it liquid\/} the mesogens lie in a smectic potential given
by, for example, a cosine modulation \cite{mcmillan71}, but are not
fixed in one trough of this potential: rather, they fluctuate back and
forth over the barriers.  In a strongly-ordered smectic state the
activation over the barriers is very rare, while in a weak smectic
state these fluctuations are quite common. The `motion' of the smectic
phase thus corresponds to the average motion of the mesogens'
centers-of-mass.

Now consider an elastomer in a smectic state and imagine displacing
the smectic phase while fixing the crosslink positions. For clarity
we focus on a side-chain liquid crystalline network, but the argument
applies to a main-chain network as well. The displacement of the smectic
phase displaces the average positions $\{\eta_i\}$ of the mesogenic
sidegroups. Since these are tethered to the polymer backbone, this costs
roughly the entropy of displacing the center-of-mass of a polymer chain
a distance $u$ while keeping the endpoints ({\it i.e.\/} crosslink
positions) fixed.  A simple calculation leads to an energy per chain
of $V_u \sim k_{\scriptscriptstyle B}T u^2 / (N a^2)$, with $N$ the
number of monomers of size $a$ between crosslinks.  In addition
to this elastic force, an individual mesogen feels a force due to the
smectic potential \cite{mcmillan71}.  At zero temperature this potential
enforces the separation of the smectic phase variable and the crosslink
displacement, but at finite temperature this barrier may be overcome
by hopping. By hopping we mean that the mesogen flips from one smectic
trough to another, in such a way as to relax the strain. This allows the
smectic phase variable to increase without bound while roughly localizing
the mesogens' centers-of-mass.

Hence a timescale $\tau^{\ast}$ separates liquid-like smectic behavior
at long times from `solid-like' smectic behavior at short times, the
latter characterized by the elastic modulus $\Lambda$. For
temperatures above the nematic-smectic transition temperature
$\tau^{\ast}$ is extremely small and we may ignore this elastic
effect. However, as the network is cooled deep into the smectic state
$\tau^{\ast}$ grows, and for all practical purposes we must include
this term. Since we are only concerned here with very weak smectic
phases in this work, we leave further discussion of the interesting
dynamics of a smectic network to another work.

\section{The model}

In this section we introduce the full free energy of the system, and
set up the replica calculation. The NA transition in a liquid is
described by the Landau-de~Gennes free energy,
\begin{equation}
2F_{\scriptscriptstyle\it NA} = \int\!d^3\!r\,\left\{
\tau|\psi({\rm\bf r})|^2 + \case12 \beta |\psi({\rm\bf r})|^4 +
g_{\perp}
|(\bbox{\nabla}_{\perp} - iq_0\delta\bbox{\hat{n}})\psi({\rm\bf r})|^2
+ g_{\parallel} |\partial_z\psi({\rm\bf r})|^2 \right\},
\end{equation}
where the minimal coupling $(\bbox{\nabla}_{\perp} -
iq_0\delta\bbox{\hat{n}})$ satisfies rotational invariance
of the director and the layered system \cite{pgdg}. In the
high-temperature nematic phase director distortions
$\delta\bbox{\hat{n}}$ are penalized by the Frank free energy,
\begin{equation}
2F_{\it fr} = \int\!d^3\!r\,\left\{
K_1 (\bbox{\nabla}\!\cdot\!\delta{\bf\hat{n}})^2
+ K_2 ({\bf\hat{n}}\!\cdot\!\bbox{\nabla}\!\times\!
\delta{\bf\hat{n}})^2
+ K_3 ({\bf\hat{n}}\!\times\!\bbox{\nabla}\!\times\!
\delta{\bf\hat{n}})^2 \right\}.
\label{eq:frank}
\end{equation}
In this work we use the one constant approximation,
$K_1=K_2=K_3\equiv K$.
The elastic energy of an underlying uniaxial solid can be written in many
ways \cite{landauelastic}, and we choose the representation
through the traceless strain tensor
$\bar{\epsilon}_{\alpha\beta}\equiv\epsilon_{\alpha\beta}-
\case13{\rm Tr}[\bbox{\epsilon}]\delta_{\alpha\beta}$, in
order to deal explicitly
with the case of a nearly incompressible network:
\begin{equation}
2 F_{\it el}
= \int\!d^3\!r \, \left\{ C_1 \bar{\epsilon}_{zz}^2 + 2 C_2
\bar{\epsilon}_{zz} {\rm Tr}[\bbox{\epsilon}]
+ C_3 \left({\rm Tr}[\bbox{\epsilon}]\right)^2
+ 2 C_4(\bar{\epsilon}_{xx}^2 + 2
\bar{\epsilon}_{xy}^2 + \bar{\epsilon}_{yy}^2) + 4 C_5
(\bar{\epsilon}_{xz}^2 + \bar{\epsilon}_{yz}^2) \right\} ,
\label{eq:felastic}
\end{equation}
where the $\hat{z}$-axis is chosen along the nematic director.
In the isotropic case, the elastic moduli $\{C_i\}$
transform to the Lam\'{e} coefficients as $C_1 = 2 \mu \, ; \  C_2 = 0
\, ; \  C_3   = \lambda + \frac{2}{3}\mu \, $ and  $ \,  C_4 = C_5 =
\mu$.  In most cases we take the limit
$C_3\rightarrow\infty$, since $\lambda/\mu\sim 10^4-10^5$ in a typical
rubber. The rubber shear moduli scale as $\mu\sim N_x
k_{\scriptscriptstyle B}T$, where $N_x\sim\ell_x^{-2}$ is the number of
crosslinks per unit volume.

The rubber--nematic free energy penalizing relative
rotations of the director $\bbox{\omega}=\bbox{\hat{n}}\times\delta
\bbox{\hat{n}}$ and elastic strain
$\bbox{\Omega}=\case12(\nabla\times{\rm\bf v})$ can be written as
\cite{pgdg80one,olmsted94}
\begin{equation}
2 F_{\it n-u} = \int\!d^3\!r \, \left\{
b_1 \left[ (\bbox{\Omega}-\bbox{\omega})\!\times\!{\bf\hat{n}}\right]^2 +
2b_2\,{\bf\hat{n}}\!\cdot\!\bbox{\epsilon}\!\cdot\!
(\bbox{\Omega}-\bbox{\omega})\!\times\!{\bf\hat{n}}  \right\},
\label{eq:n-u}
\end{equation}
where
$b_1$ and $b_2$ are proportional to, respectively,
the nematic order parameter $Q^2$ and $Q$, multiplied by
$N_xk_{\scriptscriptstyle B}T$. A positive coupling $b_2$ corresponds to a
network which favors parallel alignment between mesogen
and backbone orientations, while a negative $b_2$ favors perpendicular
alignment.

For the special case of nematic elastomers which possess an arbitrary
choice of the nematic axis (because of a spontaneously broken symmetry
of a network, formed in the isotropic state) rather than a `quenched'
axis of alignment dictated by, for example, an applied field during
crosslinking or crosslinking in the nematic state, the relationship
$C_5 - b_2^2/(4 b_1)=0$ holds \cite{golubovic89,olmsted94} and the phonons
described by $\bar{\epsilon}_{xz}$ and $\bar{\epsilon}_{yz}$ are
``soft'' ({\it i.e.\/} the corresponding phonons have fluctuation
spectrum $\langle |u(q)|^2\rangle\sim q^{-4}$ instead of the
conventional $q^{-2}$ behavior).
The deviation of the network from this curious soft case is thus
parameterized by
\begin{equation}
C_5 - {b_2^2\over 4 b_1} \equiv \Delta.
\label{eq:csoft}
\end{equation}
The presence or absence of these modes plays an essential role in
describing the NA transition. While a perfect soft
nematic elastomer ($\Delta=0$) is unlikely to be found in typical experiments
due to random stresses in the nematic state or a predetermined
alignment direction during network formation, we include the
soft case to present a complete theoretical picture.

The partition function for the system is then given by
\begin{equation}
{\cal Z\/} = \int\! {\cal D\/}\psi
{\cal D}{\rm\bf v}{\cal D} \delta\bbox{\hat{n}} \ \exp-{1\over
k_{\scriptscriptstyle B}T } \bigl\{F_{\scriptscriptstyle NA} +
F_{\it fr} + F_{\it el} +  F_{\it n-u} + F_{\scriptscriptstyle RF}
\bigr\}.
\label{eq:Z0}
\end{equation}
To find the effects of crosslinks on the NA transition
in the network and, eventually, on the nature of the smectic state,
we average over the disorder associated with these
crosslinks.
To compute quantities such as the quenched disorder-averaged free energy
and correlation functions we
use the replica trick \cite{edwardsanderson75} to write the free energy
of the system as
\begin{equation}
F/k_{\scriptscriptstyle B}T  = - \langle\log{\cal Z\/}\rangle_P =
\lim_{n\rightarrow 0} {1 - \langle{\cal Z\/}^n\rangle_P \over n},
\label{eq:replicatrick}
\end{equation}
introducing $n$ replicas of the system.
The disorder average over the distribution $P[c]$, Eq.~(\ref{eq:Pc}),
couples the replicas together into the term $F_{\it repl}$:
\begin{eqnarray}
2F_{\it repl} & = &
-\frac{\gamma^2 N_x}{k_{\scriptscriptstyle B}T}
\sum_{a,b=1}^n\int\!d^3\!r \, |\psi^a({\rm\bf r}+{\rm\bf v}^a
({\rm\bf r}))| |\psi^b({\rm\bf r}+{\rm\bf v}^b ({\rm\bf r}))|
\label{eq:fxl1} \\
&  & \ \ \  \times\, \cos\left[q_0\left(z - u^a({\rm\bf r}) +
v_z^a\left({\rm\bf r}\right) \right)\right]\,
\cos\left[q_0\left(z - u^b({\rm\bf r}) +
v_z^b({\rm\bf r})\right)\right]\, .
\nonumber
\end{eqnarray}
Note that $u$ fluctuations are implicitly included in the measure for
$\psi$ fluctuations. From here on we consider a uniform $\psi$ for
the purposes of understanding the mean-field behavior of the transition,
and as such will not consider fluctuations of its phase.

We next coarse-grain the system by averaging  over the period of
the smectic modulation $q_0^{-1}$ along the layer normal $z$.
This procedure requires that both the `mean-free-path'
between crosslinks $\ell_{x} \sim N_x^{-1/2}$ and typical length scale
of the relative translations $u({\rm\bf r}) -
v_z({\rm\bf r})$ be large compared to the layer spacing $2\pi q_0^{-1}$.
We find
\begin{equation}
F_{\it repl}  =
-\Gamma \sum_{a,b=1}^n\int\!d^3\!r\, |\psi^a|
|\psi^b| \cos\left[q_0\left(v^a({\rm\bf r}) - v^b({\rm\bf r})
\right)\right]  \label{eq:fxl2},
\nonumber
\end{equation}
where $\Gamma=\gamma^2 N_x/(2k_{\scriptscriptstyle B}T)$. The free energy
of the system is now given by Eq.~(\ref{eq:replicatrick}), with $n$
replicated copies of the partition function Eq.~(\ref{eq:Z0}), with
$F_{\scriptscriptstyle \mit RF}$ replaced by $F_{\mit repl}$.  Note that
the similar form of the replica Hamiltonian, containing a translationally
invariant cosine of fluctuating field $(v^a - v^b)$, appears in the problem
of random flux pinning in superconductors \cite{ledoussal95}.

\section{Mean Field Phase Behavior}

To see the effect of the crosslinks on the NA transition in the network
we integrate out the strain and director fluctuations and examine the
stability of the resulting effective theory for a uniform $|\psi|$.
This procedure is valid when the penetration length $\lambda$ for
director twist into the smectic is much smaller than the smectic
coherence length $\xi=(g_{\perp}/\tau)^{1/2}$ (this situation is often
referred to as the type-I smectic-A, by analogy with superconductors)
\cite{hlm}.  For an elastic network, nematic fluctuations have a
``mass'' $m \approx q_0^2 g_{\perp} |\psi|^2 + b_1 \Delta/C_5 $, where
the first term is the mass of due to smectic fluctuations and the
second term, $\sim N_x k_{\scriptscriptstyle B}T $, is due to the
coupling to the elastic network. As the transition is approached we
have $|\psi|^2\sim \beta/\tau$ and the twist penetration length is
given by $\lambda=(K/m)^{1/2}$, so that
\begin{equation}
{\lambda\over\xi}=\left({\tau K\over \tau q_0^2 g_{\perp}^2 +
g_{\perp}\Delta }\right)^{1/2}.
\end{equation}
This Ginzburg parameter vanishes as the critical point is approached
($\tau\rightarrow 0$), so that the condition for a type-I smectic
($\lambda/\xi\ll 1$) is trivially satisfied and the mean-field
treatment here is exact.  [Note that for the perfect soft system,
$\Delta=0$, the system may or may not be type-II, as with ordinary
smectics.] Elastic fluctuations enter the theory at Gaussian level and
may be integrated out at a mean-field level. A self-consistent check on
the neglect of these fluctuations is to ensure the smectic correlation
length at the fluctuation-induced first order transition is of order or
larger than both the mesh size (to satisfy the coarse-graining
procedure) and the twist penetration length.

To perform the calculations we choose a convenient geometry (cf.
\cite{tclchim83}). We decompose the displacement into
\begin{equation}
{\rm\bf v}=(v_z,v_{\perp},v_t),
\label{eq:geometry}
\end{equation}
where $v_z$ is parallel to the nematic director; $v_{\perp}$ is normal
to the director and belongs to the plane defined by the director and
the  wavevector~${\rm\bf q}$ in Fourier space; and $v_t$ is in the
direction defined by ${\rm\bf q}\times\bbox{\hat{n}}$. Similarly,
$\delta\bbox{\hat{n}} =(\delta\hat{n}_t,\delta\hat{n}_{\perp})$.
Integrating out the director fluctuations is straightforward since they
are not involved in the random coupling, and we find
\begin{equation}
\langle{\cal Z\/}^n\rangle_P  =  \int\!\prod_{a=1}^n
{\cal D\/}\psi^a{\cal D}{\rm\bf v}^a \exp-{1\over k_{\scriptscriptstyle B}T}
\left\{ \sum_{a} \left[F_1\left(\psi^a\right)
+  \tilde{F}_{\it el}\left({\rm\bf v}^a,\psi^a\right)\right]
+ F_{\it repl}\right\}. \label{eq:Zrep1}
\end{equation}
The effective smectic free energy
density is renormalized to
\begin{eqnarray}
{F_1(\psi)\over V} =
\case12\tau|\psi|^2 + \case14 \beta  |\psi|^4 +
k_{\scriptscriptstyle B}T \int_{\bf q}
\,\log\Bigl[ b_1 + D_0(\psi) + Kq^2 \Bigr] \ ,  \label{eq:Fdet1}
\end{eqnarray}
where $V$ is the system volume and we have used the one-constant
approximation for the bare Frank
elastic constants
Here and below, $\int_{\bf q}\equiv \int d^3q/(2\pi)^3$.
The parameter $D_0= |\psi|^2 g_{\perp} q_0^2$ is proportional to the
square of the smectic order parameter, and
the logarithm in Eq.~(\ref{eq:Fdet1}) is the determinant of the
quadratic form from integrating the director fluctuations. The factor
$D_0(\psi)$ in this term
gives rise to the Halperin-Lubensky-Ma effect in a mean-field type-I
system. That is, when the director is not coupled to
the network ($b_1=0$), director fluctuations are  massless and the
$q$-integration renormalizes the transition
temperature  $T_{\scriptscriptstyle NA}$, and yields a negative cubic term
which induces a first  order transition \cite{hlm}. However, massive
director fluctuations ($b_1\neq 0$) imply only even powers of
$|\psi|$ in the expansion of the integral, destroying the HLM effect.

Eliminating $\delta\bbox{\hat{n}}$
also renormalizes the rubber elasticity, yielding $\tilde{F}_{\it el}$,
which is a quadratic form in the phonon variable~${\rm\bf v}$. Together
with $F_{\mit repl}$, the effective energy governing the elastic
degrees of freedom is given by
\begin{equation}
{\cal H} =
 \sum_a \tilde{F}_{\it el} + F_{\it repl} =
\frac{1}{2}\sum_{a}\int_{\bf q}
{\bf v}^{a}({\rm\bf q})\!\cdot\!{\bf M}(q,\psi)\!\cdot
\!{\bf v}^{a}({\rm\bf -q}) - \Gamma
\sum_{a,b}\int\!d^3\!r  |\psi^a||\psi^b|\cos q_0(v_z^{a}-v_z^{b}),
\label{eq:qualreplH}
\end{equation}
where the kernel matrix ${\bf M}(q,\psi)$ in $\tilde{F}_{\mit el}$
is given in the Appendix.

The next step is to integrate out the phonon variable
${\rm\bf v}({\rm\bf r})$ and extract the effective free energy as a function
only of $\psi$.  Since ${\cal H\/}$ is non-Gaussian this is a
non-trivial step. We perform this integration by applying the Gaussian
Variational Method (GVM) as introduced by Mezard and Parisi
\cite{mezardparisi91} for random systems with translationally invariant
(in replica space) Hamiltonians, like our $\cos [v^a - v^b]$.  In
performing this integration we assume replica symmetry for the phonon
fluctuations, which holds for small enough $\psi$ and is thus justified
for an analysis of a continuous transition. We also assume replica
symmetry in the value of $\psi$ which extremizes the partition
function, so that we present below an effective free energy of $\psi$
which does not involve replicas and which we believe describes the NA
transition in a random network.

The integration of ${\rm\bf v}$ is sketched in the Appendix, and yields two
distinct theories, depending on whether or not the system exhibits `soft'
elasticity ($\Delta=0$). The effective free energies are:
\begin{eqnarray}
{F_{\psi}\over V} &=& \left\{
\begin{array}{lr}
\case12  {\tau}_0  |\psi|^2  - \lambda_0
|\psi|^3 +  \case14{\beta}_0  |\psi|^4 & \qquad (\Delta=0) \\
\noalign{\vskip6pt}
\case12{\tau}_{\Delta}|\psi|^2 + \case14 {\beta}_{\Delta}  |\psi|^4 &
 (\Delta\neq0)
\end{array}\right.
\label{eq:Flandboth}
\end{eqnarray}
which can also be presented as a single crossover expression covering both
cases:
\begin{eqnarray}
 {F_{\psi}\over V} =
\case12  \tilde{\tau}(\Delta)  |\psi|^2
+  \tilde{\lambda}(\Delta,\psi) +
\case14\tilde{\beta}(\Delta) |\psi|^4
\label{eq:Flandcrossover}
\end{eqnarray}
The free energies above comprise one of the primary results of this
work.  In Eq.~(\ref{eq:Flandcrossover}) the coefficients
$\tilde{\tau}(\Delta)$ and $\tilde{\beta}(\Delta)$ are continuous
analytic functions of $\Delta$, while $\tilde{\lambda}(\Delta,\psi)$ is
a non-analytic function which gives the cubic term for $\Delta=0$. This
term is responsible for the qualitative results we find. In this work
we give only the detailed expression for the cubic crossover function
$\tilde{\lambda}(\Delta,\psi)$, since this term alone determines the
qualitative aspects of the transition.

\underline{\sl Soft nematic elastomers\/} ($\Delta=0$)---The
renormalized coefficient ${\tau}_0$ has the form $\tilde{\tau}_0
(T-T_{\scriptscriptstyle NA})$, with the new transition
temperature shifted by the effects of rubber elasticity and random
pinning (\ref{eq:fxl2}). Besides this obvious effect,
the new feature of this soft regime is the
cubic term, which restores the HLM effect. This term is given by
\begin{equation}
\lambda_{0} = - {1\over 12\pi}\left({q_0^2 g_{\perp}\over K}\right)^{3/2},
\end{equation}
and is {\it precisely\/} the cubic term given by Halperin-Lubensky-Ma
\cite{hlm}. Hence, while the coupling of elasticity to the director (or
gauge) field destroys the HLM effect, the effect is restored if the
coupling preserves the set of Goldstone modes present in the nematic
state, an intuitively pleasing situation.

\underline{\sl Conventional nematic elastomers\/}
($\Delta\neq 0$)---In a network formed in the nematic state
there are no soft elastic phonons and no Goldstone modes, and the
effective Landau-de Gennes free energy $F_{\psi}$ contains renormalized
quadratic and quartic coefficients. There are numerous contributions to the
quartic terms, but the qualitative nature of the renormalized quartic
follows from examining the behavior of the non-analytic crossover
function, given by
\begin{eqnarray}
\tilde{\lambda({\Delta},\psi)} & = &  - {1\over 12\pi K^{3/2}}\left[
g_{\perp}q_0^2|\psi|^2 + {4 \Delta b_1^2 \over 4 \Delta b_1 + (b_1 + b_2)^2}
\right]^{3/2}    \label{eq:LamD} \\
& \simeq & -{1\over 12 \pi K^{3/2}} \left[ {\cal O\/}(1,
|\psi|^2) + \case38 g_{\perp}^2 q_0^4{b_1 + b_2\over b_1} {|\psi|^4\over
\Delta^{1/2}} + \ldots\right],\quad
\left({g_{\perp} q_0^2 |\psi|^2\over b_1} <
{4\Delta b_1\over (b_1 + b_2)^2}\ll 1\right), \label{eq:LamDapprox}
\end{eqnarray}
where the first inequality defines the validity of expanding the $3/2$
power, and the second inequality allows us to simplify the results for a
nearly-soft system ($\Delta\simeq 0$).

Hence, the non-analytic term yields a
negative quartic coefficient when expanded for small $|\psi|$ at non-zero
$\Delta$. For small enough $\Delta$ this term becomes very large and the
expansion is only relevant for extremely small $|\psi|$. At the point where the
expansion becomes invalid the quartic term must be replaced by a cubic term,
and
a tricritical point results. Alternatively, one may examine the free
energy above, Eq.~(\ref{eq:Flandcrossover}), to find a tricritical point
at $\Delta_{\ast}$ given (for small $\Delta$) by
\begin{equation}
\beta_{\Delta_{\ast}}\approx \beta
-{1\over 32\pi}\left({g_{\perp} q_0^2\over \Delta_{\ast} K^3}\right)^{1/2}
{b_1 + b_2\over b_1},
\end{equation}
where $\beta$ is the bare quartic coefficient. The NA transition is of
second (first) order for positive (negative) $\beta_{\Delta}$.

Thus, for small enough $\Delta$ the transition is expected to be of
first order.  This correction is most important in those systems which
are already close to a tricritical point, $\beta\agt0$ [typically those
with a relatively small nematic range, $T_{\scriptscriptstyle IN} -
T_{\scriptscriptstyle NA}$, where $T_{\scriptscriptstyle IN}$ is the
isotropic-nematic transition temperature \cite{mcmillan71,pgdg}]. We have
checked numerically that, for a wide range of realistic elastic constants,
the correction to $\beta$ ranges up to $0.1\,k_{\scriptscriptstyle B}T
q_c^3\sim\beta$, where $q_c$ is the microscopic (large wavenumber) cutoff
that defines the coarse-graining of the theory \cite{pgdg}. As noted in
the Appendix, the {\sl only\/} effect of disorder is to slightly increase
the transition temperature, which is sensible because crosslink sites
locally enhance the smectic order.  Our effective free energy assumes
a replica-symmetric ground state. In the Appendix we note that the
replica-symmetric solution is stable for $|\psi|=0$, while in the smectic
state there is a crossover at small $|\psi|$ to a replica--non-symmetric
state, which in this context refers to the effect of the smectic order
on the background phonon spectrum. An analysis of the further effects of
disorder on the low temperature state is beyond this scope of this work.

\section{Conclusion}
\subsection{Experimental consequences}

We conclude by first discussing some experimental consequences and
signatures of this work.  The primary qualitative prediction concerns the
order of the transition. The preparation of networks of varying degrees
of `softness' (which may be controlled by, for example, changing the
degree of order which is frozen into the smectic state) allow for tests
of the predicted tricritical behavior between first- and second-order
nematic--smectic-{\sl A\/} transitions, which may be probed by, for
example, specific heat experiments.

It would also be interesting to test the crossover to a `glassy' phase
characterized by replica-symmetry breaking. X-ray scattering to probe
the smectic density wave should yield this information, in the form
of a crossover from a Landau-Peierls' form to a structure factor with
temperature-{\sl independent\/} logarithmic correlations, similar to the
vortex glass considered by Korshunov \cite{korshunov93} and Giamarchi
and Le~Doussal \cite{giamarchi95}.

A cornerstone of our treatment involves the coupling between the smectic
phase variable and the elastic strain, in Eq.~(\ref{eq:new}), which we have
asserted to be a dynamic effect. A signature of this would be
a characteristic timescale $\tau^{\ast}$ associated with hopping of
mesogens over the smectic barrier. In strong smectic phases this
timescale should be essentially infinite so that we may use
Eq.~(\ref{eq:new}), but it should be present in weak smectics and emerge
in measurements of the complex rheological response function.
\subsection{Critique and outlook}
In this work we have examined the onset of the nematic--smectic-A
transition in a smectic elastomer, and delineated two types of behavior
(summarized in the free energies, Eqs.~\ref{eq:Flandboth}). Soft
nematic elastomers should display a first order transition, due to the
HLM effect induced by the spectrum of Goldstone modes which are a
combination of director and strain fluctuations. Conventional `hard'
elastomers, on the other hand, should have a continuous transition,
with a tricritical point as the soft limit is approached.

In performing our calculations we have assumed the system is in the
type-I limit, so that a mean field treatment is valid. Since director
fluctuations have an additional mass due to the coupling to the elastic
network, we believe this limit is safe for smectic elastomers.  We have
treated disorder within the replica formalism and found that, at
the level of the onset of the transition, the only effect is to
stabilize the smectic state. A preliminary analysis suggests that the
effects of disorder are certainly important deep in the smectic state.

Hence a natural extension of this work concerns the nature of the low
temperature phase. The effects of disorder are two-fold: (1)
Translational symmetry is broken and the smectic phase is locally
pinned to the disorder, which should destroy the Landau-Peierls
transition in favor of true long-range order of the smectic density
wave; and (2) disorder can be strong enough to destroy the long-range
order itself. Which, if either, of these effects wins out is an
interesting question, analogous to the effects of disorder on other
systems with continuous symmetry, such as the XY-model or flux lattices
in superconductors.

A second, more speculative, result of our work is the suggestion that
the coupling proposed previously to describe the energy penalty for
sliding the smectic phase relative to the crosslinks,
Eq.~(\ref{eq:new}), is actually present only when one considers
sufficiently short timescales (which, practically, may still be of
order of years). This leads to predictions of rheological effects, as
well as possible glassy phases to `freeze' this term in, which we leave
for the future. This interesting suggestion is based on the notion that
the smectic layers are actually `phantom' layers, and their motion need
not correspond with center-of-mass motion of the mesogens.

We are grateful to D.~Khmelnitskii, T.~Lubensky, and M.~Warner for
helpful conversations. This work has been supported by  Unilever-PLC (EMT) and
the EPSRC (PDO).

\appendix
\section{Integration of fluctuation modes}
In this Appendix we sketch the steps to integrate the network phonon
field ${\rm\bf v}({\rm\bf r})$ from
Eq.~(\ref{eq:Zrep1}). Recall first our coordinate system
${\rm\bf v}=(v_z,v_{\perp},v_t)$, as given in Eq.~(\ref{eq:geometry}).
We begin with the energy governing the elastic degrees
of freedom, Eq.~(\ref{eq:qualreplH}), after integrating out the director
degrees of freedom:
\begin{equation}
{\cal H} = \frac{1}{2}\sum_{a}\int_{\bf q}  \
{\bf v}^{a}({\rm\bf q})\!\cdot\!{\bf M}(q)\!\cdot
\!{\bf v}^{a}(-{\rm\bf q})  - \Gamma
\sum_{a,b}\int\!d^3\!r  |\psi^a||\psi^b| \cos \,q_0\, {\bf\hat{z}}\!\cdot\!
({\rm\bf v}^{a}-{\rm\bf v}^{b}),
\label{eq:qualreplH2}
\end{equation}
where the matrix ${\rm\bf M}\left({\rm\bf q},\psi^a\right)$ is given by
\begin{equation}
{1\over g_n(q)}\left(
\begin{array}{ccc}
C_{zz} q_z^2 g_n(q) + C_{51}'  q_{\perp}^2 + K\alpha_1 q^2 q_{\perp}^2 &
q_z q_{\perp} (C_{\perp z} + K\alpha_{\perp} q^2) & 0 \\
\noalign{\vskip5pt}
q_z q_{\perp} (C_{\perp z} + K\alpha_{\perp} q^2) &
C_{\perp\perp} q_{\perp}^2 g_n(q) + C_{52}' q_z^2 +
K\alpha_2 q^2 q_z^2 & 0 \\
\noalign{\vskip5pt}
0 & 0 & C_4 q_{\perp}^2 g_n(q) + C_{52}' q_z^2 + K\alpha_2 q^2 q_z^2
\end{array}
\right),
\label{eq:matrix}
\end{equation}
where $g_n(q)=b_1 + D_0(\psi) + K q^2$ and the new renormalized elastic moduli
are
\begin{equation}
\begin{array}{rcl@{\qquad}rcl}
C_{\perp\perp} & = & C_3 + \case19C_1 - \case23C_2 +\case{10}9 C_4 &
C_{zz} & = & C_3 + \case49C_1 + \case43C_2 + \case49 C_4 \\
\noalign{\vskip4pt}
C_{\perp z} & = &  (b_1 + D_0) \alpha_{\perp} + (b_1^2 - b_2^2) / 4 &
\alpha_1 &=& \Delta + (b_1 - b_2)^2/(4 b_1) \\
\noalign{\vskip4pt}
\alpha_2 &=& \Delta + (b_1 + b_2)^2/(4 b_1) &
\alpha_{\perp} &=& C_3 - \case29C_1 + \case13C_2 - \case29 C_4 + C_5  \\
\noalign{\vskip4pt}
C_{51}' & = & \Delta b_1 + D_0 \alpha_1  &
C_{52}' & = & \Delta b_1 + D_0 \alpha_2.
\end{array}
\end{equation}
Recall that $\Delta = C_5-b_2^2/(4 b_1)$ is the
deviation from an ideal soft nematic elastomer \cite{golubovic89,olmsted94}.

We first rescale ${\rm\bf v}$, pulling out the determinant of ${\bf M}({\bf
q})$, which leaves the following replica partition  function:
\begin{equation}
\langle{\cal Z\/}^n\rangle =\int\prod_a{\cal D\/}\psi^a
\exp -{1\over k_{\scriptscriptstyle B}T } \left\{\sum_a F_1(\psi^a) +
\case12\sum_a k_{\scriptscriptstyle B}T \int_{\bf q}
\log\det{\bf M}({\bf q},\psi^a)  + F_w(\psi) \right\}
\end{equation}
where
\begin{equation}
e^{\displaystyle -F_w(\psi)/k_{\scriptscriptstyle B}T} =
\int\prod_a{\cal D\/}{\rm\bf w}^a \exp -{1\over k_{\scriptscriptstyle B}T }
\left\{\case12\sum_a\int_{\bf q} |{\rm\bf w}^a({\rm\bf q})|^2
- \Gamma\sum_{a,b}\int \!d^3\!r |\psi^a||\psi^b| \cos q_0 \left[
{\bf\hat{z}}\!\cdot\!{\bf M}^{-1/2}\!\cdot
\!\left({\rm\bf w}^a - {\rm\bf w}^b\right)\right]\right\}
\end{equation}
Note that the matrix ${\bf M}({\rm\bf q})$ becomes a differential operator in
real space, with the square root defined with reference to the
operator in Fourier space.

To calculate $F_w(\psi)$ we use the Gaussian Variational Method (GVM)
\cite{edwardsmuthukumar88,mezardparisi91}. Since this method is well-known and
is essentially the Hartree approximation, we sketch the procedure and
refer the reader to references
\cite{edwardsmuthukumar88,ledoussal95,mezardparisi91} for further
discussion. We begin by assuming a trial Hamiltonian,
\begin{equation}
{\cal H\/}_t=\frac{1}{2}\sum_{a,b}\int_{\bf q}
{\bf w}^{a}({\rm\bf q})\!\cdot\!{\bf G}_{a,b}^{-1}({\rm\bf q})\!\cdot
\!{\bf w}^{b}(-{\rm\bf q}).
\end{equation}
The free energy for the ${\rm\bf w}$ integration satisfies
the inequality
\begin{equation}
F_w\leq F_{\mit var} \equiv F_t + \langle {\cal H\/} - {\cal H\/}_t \rangle_t,
\end{equation}
where the average is taken with the trial Hamiltonian ${\cal H\/}_t$.
The variational free energy $F_{\mit var}$, which  must then be minimized over
the matrix ${\bf G}_{a,b}({\rm\bf q})$, is
\begin{equation}
F_{\mit var} = -{k_{\scriptscriptstyle B}T \over 2}\left[
\int_{\bf q} \left({\rm Tr}\log {\bf G} -
{\rm Tr}\,{\bf G}\right) + n\right] -\Gamma\sum_{a,b}\int\!d^3\!r
|\psi^a| |\psi^b| \exp \left\{-\frac{1}{2} q_0^2
k_{\scriptscriptstyle B}T \int_{\bf q} B_{a,b}({\rm\bf q})\right\},
\end{equation}
where
\begin{equation}
B_{a,b}({\rm\bf q}) = {\bf\hat{z}}_m({\rm\bf q})\!\cdot\!\left[{\bf G}_{a,a}
({\rm\bf q}) + {\bf G}_{b,b}({\rm\bf q}) - 2  {\bf G}_{a,b}({\rm\bf q})
\right] \!\cdot\!{\bf\hat{z}}_m({\rm\bf q}).
\end{equation}
Here we have defined ${\bf\hat{z}}_m({\rm\bf q})\equiv{\bf
M}^{-1/2}({\rm\bf q})\!\cdot\!{\bf\hat{z}}({\rm\bf q})$.

Next we calculate the self-consistency conditions $\delta F_{\mit
var}/\delta {\bf G}_{a,b}=0$. For a replica-symmetric ansatz,
${\bf G}_{a,b} = a_0\bbox{\delta}_{a,b} + {\bf a_1}$, the self-consistency
conditions yield $a_0=1$ and
\begin{eqnarray}
{\bf a_1} & = & -2\,{\bf\hat{z}}_m({\rm\bf q})\,{\bf\hat{z}}_m({\rm\bf q})
\,\Gamma\,\hat{\gamma}\,q_0^2\int\!d^3\!r |\psi^2| \\
\noalign{\vskip5pt}
\hat{\gamma} & = & \exp \left\{- q_0^2\, k_{\scriptscriptstyle B}T
\int_{\bf q} {\bf M}^{-1}_{zz}({\rm\bf q}) \right\}.
\end{eqnarray}
In obtaining this relation we have assumed that the
the smectic instability is satisfactorily described
by a replica-symmetric ansatz $|\psi^a|=|\psi|$.
Henceforth we discard free energies terms of higher
than linear order than $n$, which vanish in the $n\rightarrow 0$ limit.
Moreover, we expect that in the low temperature phase
$\langle|\psi^a|\rangle_P$ is replica-symmetric, and glassy
behavior manifests itself in broken replica symmetry of the phase
variable [{\it i.e.\/} the layer spacing $u({\rm\bf r})$] as in, for
example, the $XY$-model in a random field
\cite{cardy82,ledoussal95,giamarchi95}.

Next we identify the free energy
using Eq.~(\ref{eq:replicatrick}), and examine the
stability of the nematic phase against uniform smectic order.
Hence we examine the stability of
${F_{\psi}}=k_{\scriptscriptstyle B}T \lim_{n\rightarrow0} (1 -
\langle{\cal Z\/}^n\rangle_P)/n$.  The final result is:
\begin{equation}
{F_{\psi}\over V} = \case12\tau|\psi|^2 + \case14 \beta |\psi|^4 +
k_{\scriptscriptstyle B}T \int_{\bf q}
\,\log g_n(q) + \case12 k_{\scriptscriptstyle B}T\int_{\bf q}
\log\det\,{\rm\bf M}\left({\rm\bf q},\psi\right) -
\Gamma\,|\psi^2|\,(1-\hat{\gamma}).
\label{eq:Fpsibig}
\end{equation}

There are four $q$-integrals to perform: one ($I_1$) from $\log g_n(q)$,
which arises from director fluctuations and from the
logarithm of the prefactor of the determinant of ${\rm\bf M}$. These are
identical, and exhibit a partial cancellation [a factor of $1$ from the
director
fluctuations and a factor of $-3/2$ from the determinant of ${\rm\bf M}$];
two more integrals ($I_2$ and $I_3$) emerge from $\log\det{\rm\bf M}$, and one
from  ($I_4$) ${\rm\bf M}^{-1}_{zz}$, which enters in the exponential in
$\hat{\gamma}$.

$I_1$ may be performed exactly and expanded to yield even powers of $\psi$.
We take $I_2$ to be the integral of the logarithm of the
single diagonal of ${\rm\bf M}$ involving $C_4$ (that is,
${\rm\bf M}_{tt}$). This has a contribution
non-analytic in $\psi$ for $\Delta\rightarrow 0$, since in this case
$C_{52}'\sim |\psi|^2$ and the integrand in the ${\rm\bf q}$-integral is
singular as $q_z\rightarrow 0$.  This yields the cubic term
$\lambda(\Delta,\psi)$, as well as many terms even in $\psi$. $I_3$ is then
the logarithm of the determinant of the remaining $2\times2$ sector of
${\rm\bf M}$. This is also non-analytic in $\psi$ at $\Delta=0$, but yields
terms of order $|\psi|^5, \psi^4\log|\psi|$, {\it etc\/}, which we ignore
compared to the effect of $\psi^3$ contribution (these terms can be shown to
lead to subdominant terms for small $\Delta$).

The last integral, appearing in $\hat{\gamma}=e^{I_4}$, must be handled
with care. We are interested in expanding this integral for small $\psi$
for both the soft ($\Delta=0$) and hard ($\Delta\neq 0$) cases.
For $\Delta, \psi \longrightarrow 0$ the integrand is singular
as $q_z\rightarrow 0$. This singularity yields a logarithmic
contribution to the integral, so we write
\begin{equation}
\hat{\gamma}=e^{ - q_0^2 k_{\scriptscriptstyle B}T
\int_{\bf q} {\bf M}^{-1}_{zz}({\rm\bf q})} =
|\psi|^{\eta}e^{-(I_0 + f_0(\psi))},
\end{equation}
where $I_4^0$ is a constant and $f_0(\psi)$ is an analytic function of $\psi$.
Hence the term involving $\hat{\gamma}$ in Eq.~(\ref{eq:Fpsibig}) contributes
a term in the energy proportional to $\psi^{2+2\eta}$, multiplied by an
exponentially small prefactor. We find $\eta$ to be essentially the
Caill\'e exponent
\cite{caille}, $\eta=q_0^2 k_{\scriptscriptstyle B}T \sqrt{b_1/K} /
(8 \pi C)$, where $C$ is a combination of rubber elastic constants.  Since
$\eta$ may be of order $0.5-2.0$, we ignore  this term compared to the cubic
term. If $\eta$ is sufficiently small we  may include this term in a Landau
expansion, but  we expect the exponentially small prefactor to render it
negligible. In  the case where $\Delta$ is non-zero the integral is analytic
and yields even powers of $\psi$. Therefore, the only qualitative
effect of disorder is to {\sl increase\/} the transition temperature.

The final step of the calculation is to ensure stability of the
replica-symmetric ansatz for the ${\rm\bf v}$ integral. This is done by
examining the replicon mode, identified for the GVM as \cite{mezardparisi91}
\begin{equation}
\lambda = 1 - \Gamma |\psi|^2 q_0^4 k_{\scriptscriptstyle B}T
\hat{\gamma} \int_q ({\rm\bf M}^{-1}_{zz})^2,
\end{equation}
where $\lambda$ is the eigenvalue of the most unstable fluctuation mode
about the replica-symmetric solution given by ($a_0, {\bf a_1}$). We note
that this is actually the eigenvalue of the fluctuation mode of the
$N_v\rightarrow\infty$ limit version of the theory, where $N_v$ is the
number of `color' components of the field ${\rm\bf v}$, in which limit the
GVM (or Hartree) approximation becomes the exact saddle point integral
\cite{mezardparisi91}.
While $\lambda$ is calculated in this limit, it may or may not
apply to the physical case $N_v=1$, but hopefully yields intuition about
the correct behavior.

Since $\lambda=1$ in the high-temperature phase
(because $|\psi|=0$), the replica-symmetric solution is stable as far
as understanding the onset of the smectic transition. Analysis
reveals an $L$-divergence in $\int_q ({\rm\bf M}^{-1}_{zz})^2$.
which indicates
a critical temperature (in the smectic state) at which the disorder is
relevant and we must consider replica--non-symmetric states. Here $L$ is
the system dimension. Ignoring numerical factors, this `glass' transition
temperature $T^{\ast}$ is given by the temperature very slightly below
the NA transition at which the order parameter attains the value
$\psi^{\ast}$, given by
\begin{equation}
|\psi^{\ast}| \sim \left({C^2\over\gamma^2 N_x q_0^4 L}\right)^{1\over
2+\eta},
\end{equation}
where $C\sim k_{\scriptscriptstyle B}T N_x$ is a combination of rubber
elastic constants. Since $\psi^{\ast}$ is inversely proportional to the
system dimension $L$, the window within which the replica-symmetric
solution holds is expected to be very small.  Thus, for understanding
the properties of realistic systems, one must examine the properties of
the disordered low temperature state within the framework of
replica--symmetry-breaking.



\end{document}